# Versatile silicon-waveguide supercontinuum for coherent mid-infrared spectroscopy

Nima Nader,[1*] Daniel L. Maser,[2,3] Flavio C. Cruz,[2,4] Abijith Kowligy,[2] Henry Timmers,[2] Jeff Chiles,[1] Connor Fredrick,[2,3] Daron A. Westly,[5] Sae Woo Nam,[1] Richard P. Mirin,[1] Jeffrey M. Shainline,[1] and Scott A. Diddams[2,3*]

[1]Applied Physics Division, National Institute of Standards and Technology, 325 Broadway, Boulder, Colorado 80305, USA

[2]Time and Frequency Division, National Institute of Standards and Technology, 325 Broadway, Boulder, Colorado 80305, USA

[3]Department of Physics, University of Colorado Boulder, 2000 Colorado Boulevard, Boulder, Colorado 80309, USA

[4]Instituto de Fisica Gleb Wataghin, Universidade Estadual de Campinas, Campinas, SP, 13083-859, Brazil

[5]Center for Nanoscale Science and Technology, National Institute of Standards and Technology, 100 Bureau Drive, Gaithersburg, Maryland 20899, USA

Corresponding Authors: nima.nader@nist.gov, scott.diddams@nist.gov

**Infrared spectroscopy is a powerful tool for basic and applied science. The molecular "spectral fingerprints" in the 3 μm to 20 μm region provide a means to uniquely identify molecular structure for fundamental spectroscopy, atmospheric chemistry, trace and hazardous gas detection, and biological microscopy. Driven by such applications, the**

**development of low-noise, coherent laser sources with broad, tunable coverage is a topic of great interest. Laser frequency combs possess a unique combination of precisely defined spectral lines and broad bandwidth that can enable the above-mentioned applications. Here, we leverage robust fabrication and geometrical dispersion engineering of silicon nanophotonic waveguides for coherent frequency comb generation spanning 70 THz in the mid-infrared (2.5 μm to 6.2 μm). Precise waveguide fabrication provides significant spectral broadening and engineered spectra targeted at specific mid-infrared bands. We use this coherent light source for dual-comb spectroscopy at 5 μm.**

Spectroscopy has been a primary scientific tool for studying nature, leading to seminal advances in astronomy, quantum physics, chemistry and biology. The coherent light from a laser provides a powerful spectroscopic tool with the properties of high spectral resolution, wavelength tunability, and a well-defined Gaussian beam enabling high intensity focusing and long-distance propagation. Frequency comb lasers combine the above qualities in addition to a broad spectrum of precisely defined optical lines (the "comb") that can be absolutely referenced to radio frequencies (RF) or atomic frequency standards[1-3]. This has led to a variety of new spectroscopic advances[4-13].

While frequency combs were initially developed for the visible and near-infrared spectral regions, more recent research has focused on extending their coverage to the mid-infrared (mid-IR)[14]. This spectral region is of great interest because it is where many molecules including greenhouse gases, poisonous agents, explosives, and organics show distinctive ro-vibrational absorption fingerprints[14]. The development of a practical, broadband, and low-noise mid-IR frequency comb with moderate power could dramatically improve frequency precision, sensitivity, and data acquisition rates compared to conventional techniques such as Fourier-transform-infrared (FTIR) spectroscopy. Different approaches have been demonstrated for mid-IR comb generation including optical-parametric-oscillators (OPO)[15-18], difference-frequency-generation (DFG)[19-25], microresonators[26-30], quantum cascade lasers (QCLs)[31], and supercontinuum generation[32-38]. Nonetheless, there lies an opportunity for approaches demonstrating broad bandwidth mid-IR generation with the combined attributes of robust integration, low noise, high spectral resolution and engineered spectral shape.

In the present work, we engineer supercontinuum in silicon waveguides to realize versatile, tunable, and coherent mid-IR frequency combs. The motivation to use a nanophotonic platform lies in the connection between geometric control of the waveguides and the group-velocity-dispersion (GVD), which allows the unique tailoring of the nonlinear light generation with application-defined power, spectral shape, and bandwidth. Taken together with robust lithographic fabrication, and the strong third-order-nonlinearity of Si-on-Sapphire (SoS)[36], we realize on-chip mid-IR frequency combs across 70 THz (2.5 μm to 6.2 μm) with both multi- and broad-band spectra. We also demonstrate that the supercontinuum from these waveguides adds negligible noise beyond that of the 3 μm pump, enabling the use of these devices for dual-comb spectroscopy (DCS) of carbonyl sulfide (OCS) at 5 μm. Furthermore, the simplicity and flexibility of the photonic technology introduced here is conducive to system-level integration with emerging chip-based DFG sources[39] or mid-IR passively mode-locked fiber sources[40]. This could lead to a heterogeneously integrated nanophotonics platform for simple, versatile, mid-IR sources with small footprint, low power consumption, and modest cost. The user-controlled and engineered multiband spectra would particularly benefit applications where parallel multi-comb operation is desired, such as compact point sensors for real-time *in situ* chemical synthesis monitoring, single bio-molecule near-field microscopy, and remote sensing.

**Waveguide Geometries and Dispersion Profiles**

Silicon-based integrated photonic platform provides many characteristics required for efficient mid-IR supercontinuum generation. This includes transparency to ~ 8 μm, high third order nonlinearity ($n_2 = 6 \times 10^{-18} \frac{m^2}{W}$, which is 100x larger than silica)[36], and high index of refraction (range of 3.42 to 3.45). Together, these properties lead to low intrinsic loss, high mode confinement, versatile dispersion engineering in waveguides, and enhanced nonlinear interactions. Historically, however, nonlinear Si photonics depended on Si-on-Insulator (SOI) technology in which the optical mode interacts with a $SiO_2$ cladding material[37-38]. This platform not only limits the geometrical dispersion by reducing the core-cladding index contrast, but it also restricts the mid-IR generation to wavelengths shorter than ~ 3.0 μm due to the increased absorption of the oxide cladding. For this reason, a use of alternative Si-based platforms with different cladding materials is necessary to fully realize the advantages of an Si-based nonlinear photonic platform for mid-IR applications.

We design and fabricate air-clad Si waveguides on a sapphire substrate to avoid extra losses due to cladding and substrate absorption[36]. The waveguides are tapered to wider widths for broadband input and output coupling (see Methods for fabrication details). The propagation and coupling losses are measured at the pump wavelength to be 5 dB/cm to 7 dB/cm (depending on waveguide geometry) and 6.7 dB ± 1.4 dB per facet, respectively. Two different waveguide cross-sections are used for engineered dispersion profiles to provide anomalous GVD at the pump wavelength of 3.06 μm. Strip waveguides with rectangular cross-section of 600 nm height and different waveguide widths from 2.2 μm to 3.2 μm are fabricated as the first group of devices. A scanning-electron-micrograph (SEM) of a device cross-section overlaid with a simulation of the input mode, and a schematic diagram of the waveguides are presented in Figures 1 (a) and (b), respectively. These devices have anomalous GVD profiles (Fig. 1 (c)) that make them suitable for supercontinuum generation in which most of the optical power is transferred within 30 THz of the pump.

For efficient light generation above 4.2 µm (> 30 THz away from the pump), a different approach is taken to provide an engineered zero crossing of the dispersion at long wavelengths. A second group of waveguides, called "notch waveguides" are designed and utilized for this purpose. These waveguides have a rectangular cross section with a shallow-etched notch on top (Figs. 1 (d) and (e)). In our designs the waveguide width, $W_{wg}$, and notch depth, nD, are fixed at 3.45 µm and 300 nm respectively. Two notch widths, $W_n$, of 380 nm and 530 nm are fabricated where, as shown in Fig. 1 (f), the zero-dispersion wavelength is tuned with the notch position, nP, relative to the center of the waveguide.

These notch waveguides facilitate dispersive wave generation[41-44] where a significant portion of the pump energy, as high as 40% depending on the pump pulse duration[44], is transferred into a spectral peak with its center wavelength in the normal dispersion regime. The exact wavelength of the dispersive wave depends on the pump parameters and waveguide geometry and can be tailored using the phase matching condition[45] defined by $\beta_{DW} - \beta_P - 1/v_g(\omega_{DW} - \omega_P) = 1/2\gamma P_p$. In this expression, $\beta_{DW}$ and $\beta_P$ are the propagation constants of the waveguide at the dispersive wave and pump wavelengths, $v_g$ is the group velocity at pump, $\omega_{DW}$ and $\omega_p$ are the dispersive wave and pump angular frequencies, $P_p$ is the pump peak power, $\gamma$ is the waveguide nonlinearity and is given by $\gamma = \frac{\omega_P n_2}{c A_{eff}}$ with c being the speed of light, $n_2$ is the nonlinear index of Si at the pump wavelength, and $A_{eff}$ the effective area of the waveguide.

Strip and notch waveguides can both be designed to provide flat, anomalous GVD with values close to zero in a broad wavelength bandwidth to generate octave-span continuum for applications where broad comb coverage is needed. Here, in Fig. 1 (g) we present two designs that facilitate this goal. The first waveguide has a strip cross-section with $W_{wg}$ = 3.06 μm (dark blue curve), while the second design benefits from the notch cross-section, with $W_n$ = 530 nm and nP = 1 μm, to provide a flat GVD profile over a broader bandwidth (orange curve).

**Supercontinuum Generation**

The validity of the dispersion designs and their usefulness for broadband supercontinuum spectra is verified experimentally. We designed and built a laser based on a 1550 nm Erbium-fiber oscillator performing DFG in a periodically-poled lithium niobate (PPLN)[19] to generate 3.06 μm pump light with 100 fs, 1 nJ pulses (refer to Methods). The mid-IR pump beam is free-space coupled to the $TE_0$ mode of the waveguides, with 0.12 nJ pulse energy in the device (refer to Methods). We monitored the output of the devices with a FTIR spectrometer to record the supercontinuum spectra. We show that the strip waveguides designed in Fig. 1 (c) can be utilized for light generation in multiple bands covering the spectral range from 3 μm to 4.25 μm depending on the width of the waveguide (Fig. 2 (a)). The notch waveguides are also tested, and it is shown in Fig. 2 (b) that these can generate long-wavelength dispersive waves, pushing the light to wavelengths as high as 5.7 μm depending on the notch position. Figure 2 (c) presents the measured output spectra of the two waveguides (strip and notch) designed for broadband, octave-span supercontinuum generation. Figure 1 (g) shows that the strip waveguide provides a broad spectrum from 2.3 μm up to 5 μm where the generated spectrum drops due to the highly dispersive GVD values (> 100 ps/km·nm) at longer wavelengths. The notch waveguide generates light at wavelengths from 2.5 μm to 6.2 μm due to its extremely flat GVD profile. Increasing

GVD values and absorption of the sapphire substrate limit the generation at wavelengths above 6 µm. The output mid-IR powers of the waveguides at the shaded regions in Fig. 2 (a) and (b) are independently measured to be tens of microwatts. The spectral shape, bandwidth, and power of the shaded regions are of great importance since they can be filtered and used for dual-comb spectroscopy applications. We also noted that although 3 µm to 5 µm mid-IR light generation is possible through DFG directly from PPLN[23-24], the Si waveguide platform offers some distinct advantages. The Si-photonics platform not only enables extremely simple wavelength tunability through waveguide geometrical engineering, it also satisfies an important spectroscopic requirement of broad bandwidth (> 10 THz), and flat optical spectrum (< 10 dB spectral variation). Moreover, this platform enables having hundreds of devices with different GVD profiles on a single 1 $cm^2$ device area size.

**Comparison with model**

We model the supercontinuum generation by solving the generalized nonlinear Schrödinger equation (gNLSE) using a split-step Fourier method. The gNLSE solver includes waveguide dispersion, nonlinear phase shift, linear losses of the waveguides, three-photon-absorption (3PA), as well as 3PA-induced free carrier absorption (FCA) and dispersion (FCD) in Si according to the derivation explained in Methods. To achieve accurate results, we implement in the calculations the measured 3.06 µm pump parameters along with the measured coupling and propagation losses of the individual devices. Figures 2 (d), (e), and (f) show the calculated output spectra of the 1 cm long ridge waveguides, notch waveguides, and devices with octave-span broadening, respectively. The evolutions of the 3.06 µm pulse propagating along different waveguides are also presented in Figs. 2 (g,) (h), and (i). The pulse evolution plots correspond to the broadest spectrum in Figs. 2 (d), (e) and (f), respectively.

The pulse evolution simulations reveal that the soliton fission occurs 1 mm into the Si waveguide. While the spectral broadening in strip waveguides is mostly dominated by the soliton fission and soliton dynamics, in notch waveguides the dispersive wave generation in the normal-GVD regime is responsible for the long wavelength spectral peak as predicted by the phase matching condition explained earlier. Our use of ultrashort 100 fs pulses favors the abovementioned nonlinear processes along with self-phase modulation as the main sources of supercontinuum generation. As verified experimentally and detailed below, these processes maintain the coherence of the original pump pulse[37], in contrast to experiments with long picosecond pulses that result in background noise amplification and pulse interaction with the generated 3PA-induced free carriers[46]. We also note that the measured and simulated spectra in Fig. 2 (a) to (f) are plotted on a linear scale to emphasize the spectral flatness of the generated mid-IR bands for DCS applications.

**Noise properties**

In many spectroscopy experiments, amplitude noise present in the light source limits the achievable precision[4]. Thus, we characterize the relative-intensity-noise (RIN) for different mid-IR bands at the output of the waveguides (refer to Methods for details of the measurement). The data is taken for noise bandwidth of 100 Hz to 10 MHz and presented in Fig. 3. The intensity noise of the waveguide outputs closely follow that of the 3.06 μm pump for frequencies above 10 kHz. At less than 10 kHz, however, there is ~ 15 dB added noise at the waveguide outputs which is attributed primarily to mechanical stability and vibrations that perturb the coupling into these micron-scale waveguides. We also note a noise peak at ~ 2 MHz coming from the pump source. Significantly, the measured levels of RIN do not pose a significant limitation in our dual-comb experiments, and should be possible to reduce with a still lower-noise pump source.

**Dual-comb spectroscopy at 5 µm**

The low-noise nature of our waveguide-generated mid-IR light along with its spectral flatness in the bands of interest opens numerous applications for dual-comb spectroscopy. One candidate for such a demonstration is carbonyl sulfide (OCS) with ro-vibrational lines from 4.7 µm to 5 µm[47]. We place an OCS gas cell in the beam path of a dual-comb system centered at 4.85 µm with 250 nm full-width-half-maximum (FWHM) bandwidth. The setup diagram of the dual-comb system is presented in Fig. 4 (a) in which a Si waveguide generates a dispersive wave at 4.85 µm in one of the arms, called "comb 1", a spectrally overlapped mid-IR light is generated directly from a PPLN crystal in the second arm, "comb 2". The spectra of the Si-based and PPLN-based combs are presented in Figs. 4 (b) and (c), respectively. The Si waveguide generates 10 µW in the presented band while the PPLN crystal generates ~ 500 µW.

The pump source of the combs originates via DFG. Therefore, the mid-IR combs are offset-free, meaning that only stabilization of the repetition rates is sufficient to stabilize the combs. We lock and stabilize the $\Delta f_{rep}$ of the dual-comb system using a microwave circuit to enable recording and averaging of the dual-comb heterodyne signal. This was performed based on an RF circuit operating at the 100$^{th}$ harmonic of the repetition rates of the two combs[48]. The diagram of this circuit is presented in Fig. 4 (a) with detailed explanation in the Methods section. While this scheme has the benefit of operating with established microwave circuitry in the 10 GHz region, it is nonetheless challenging to achieve sub-half-cycle relative stability at the 600,000$^{th}$ harmonic of the repetition rate near 60 THz.

In-loop analysis of the locking noise, representing the best-case scenario for our stabilization, shows a total accumulated timing jitter of 24 fs, integrated from 10 Hz to 10 MHz. This is ~1.6x the optical cycle of 5 µm light, or equivalently 3.2π radians of phase noise. However, most of this timing jitter is accumulated between sampling rates of 1 kHz to 10 kHz. Above 10 kHz, the accumulated jitter is below 10 fs, achieving close to half-cycle relative stability. Therefore, if we choose a sampling window below 100 μs, averaging over consecutive interferogram measurements will be possible. In the experiment, we used a 20 μs acquisition window. While this is a promising result, a larger time window of $1/\Delta f_{rep} = 1/512$ Hz = 1.95 ms is required for obtaining a comb-line resolved spectrum. Achieving a sub-half-cycle relative stabilization at 5 μm (60 THz) in this full window should be possible with improved microwave signal-to-noise and higher servo bandwidth.

Figure 5 presents the spectroscopy results with the OCS gas sample in the beam path. The heterodyne beat of the dual-combs is recorded with an oscilloscope. Figure 5 (a) presents 11 successive interferograms in a 20 ms time window, with each signal being separated from the neighboring beat by 1.68 ms ($1/\Delta f_{rep}$). In Fig. 5 (b), the $\Delta f_{rep}$ is locked at 512 Hz and the acquisition window is reduced to 20 μs, of which a zoomed-in to 13 μs window is plotted for a single-shot recording with the light-blue curve. Multiple interferograms are further averaged on the oscilloscope, to increase the signal-to-noise-ratio (SNR), as shown for 16384 averages with the red curve in Fig. 5 (b). Triggering as employed here effectively aligns the center burst and removes slow jitter in $\Delta f_{rep}$. Our RF-locking scheme additionally reduces timing jitter in the 20 μs window such that we see stable interferograms with SNR that improves with averaging.

Figure 5 (c) represents the normalized optical spectrum of the dual-comb system in the presence of OCS. This is calculated from the Fourier transform of the measured interferograms with conversion of the RF frequencies to the optical domain using the values of $f_{rep}$ and $\Delta f_{rep}$ (ref. 4). Our results are also compared with that of the HITRAN 2012 database which is plotted with the filled grey curve. It is shown that our measurement is in good agreement with the database when the HITRAN spectrum is processed to match our 10 GHz spectral resolution (defined by the 20 μs measurement window). With improved frequency stabilization or more sophisticated phase correction[9], our system should ultimately provide resolution at the 100 MHz comb-tooth spacing.

**Discussion and outlook**

We presented SoS waveguides as a nonlinear photonic platform for mid-IR comb generation and spectroscopy. Through dispersion engineering, we efficiently broadcast coherent frequency comb spectra across 70 THz of bandwidth in the mid-IR with power that can be engineered on either broad or narrow spectral windows. This capability provides the means to tailor the optical spectrum for detection and analysis of specific chemical compounds, or to distinguish between multiple species. The intensity noise performance and spectral flatness of the generated light enabled us to demonstrate preliminary dual-comb spectroscopy of OCS in the 5 μm spectral region. By employing suspended Si waveguides[49] or materials such as GaP and GaAs[15, 25, 50], the approaches demonstrated here could be engineered and extended to provide spectral coverage beyond 10 μm.

A strength of this nanophotonics platform lies in the simple few-layer fabrication process along with their versatile performance and small chip footprint of 1 $cm^2$, housing hundreds of devices. This platform provides many opportunities for system-level integration with a wide range of mode-locked femtosecond-pulsed mid-IR sources. System integration enables spectral tailoring of such sources to cover the span of mid-IR regime from 3 μm to 10 μm in a flexible, controlled manner. For example, the emerging femtosecond-pulsed $Er^{3+}$-doped ZBLAN mid-IR fiber lasers[51] can be an interesting option to replace our free-space DFG-based pump source.

Alternatively, our nanophotonic chip can be integrated with a waveguide DFG pump to comprise a powerful nanophotonic system for mid-IR generation and spectral tailoring[39].

Additionally, power requirements for the pump sources can be reduced through utilization of nanophotonic devices. These reductions arise from high on-chip optical intensity, better waveguide mode matching, and GVD engineering when DFG is performed on the chip. Such a chip-based coherent infrared light source will expand applications in a range of laboratory spectroscopy, lab-on-a-chip diagnostics, and scanning probe microscopy.

**Methods**

**Waveguide Fabrication**

We start with a 600 nm thick crystalline Si layer epi-grown on an R-plane crystalline sapphire substrate. Shallow etched notches are patterned in a positive tone photoresist (SPR660) by 365 nm *i*-line step lithography. The resist is then reflowed at 150 °C for 5 minutes to reduce the sidewall roughness. These patterns are transferred to the Si layer by reactive ion etching with HBr chemistry. This etch step is timed to achieve the etch depth of 300 nm. After resist removal and wafer cleaning, the same step-lithography process is repeated to pattern the waveguides. This proceeds through the entire 600 nm Si device layer. After waveguide patterning, a 3 μm thick protective $SiO_2$ cladding layer is deposited with a plasma-enhanced chemical vapor deposition (PECVD) tool. After dicing into chips, the two end-facets are polished to achieve smooth waveguide facets for free-space coupling. The protective cladding layer is removed with buffered-oxide-etchant to achieve air-clad waveguides.

**Pump source, waveguide coupling and supercontinuum measurement**

The 3.06 μm frequency comb used to pump the Si waveguides is difference-frequency generated in a PPLN crystal. The process starts with a 100 MHz Er:fiber oscillator to generate a mode-locked, 70 fs, 1550 nm frequency comb with 30 mW average power. This light is then divided in a 50/50 fiber beamsplitter into two arms, namely pump and signal branches for the DFG process. In the pump branch, the beam is amplified in an Erbium-doped-fiber-amplifier (EDFA) and is sent through a highly-nonlinear-fiber (HNLF) with zero dispersion wavelength of ~ 1.3 µm to generate a dispersive wave at 1.03 µm. This light was then used to seed a ytterbium-doped fiber amplifier and then passed through free-space grating couplers to generate 160 fs pulses with 1 W average power. In the signal branch, the 1550 nm light is amplified in a second EDFA to 200 mW average power and then enters free space. The two arms are then combined and temporally overlapped before entering a 1 mm long PPLN crystal to generate the 3.06 µm idler beam (120 nm of optical bandwidth at the 3 dB level) with 1 nJ, 100 fs pulses (100 mW average power). A detailed diagram of this source is presented in Supplementary Section III. This TE-polarized 3.06 µm pump light is coupled into the waveguides using a chalcogenide glass aspheric lens (anti-reflection coated in 3 µm to 5 µm region) with numerical aperture (NA) of 0.56 for supercontinuum generation. After the free-space propagation of the 3 µm pump and multiple reflections and beam-taps, the average power incident on the waveguide is ~ 60 mW, with ~ 12 mW entering the device after the 6.7 dB/facet coupling loss. The waveguide alignment is optimized by imaging the output mode of the devices using a mid-IR InSb camera. After alignment optimization, the supercontinuum spectra of the waveguides are coupled out of the waveguides using an identical lens and recorded with a FTIR spectrometer. We note that due to the broad output of the waveguides and achromatic aberration of the out-coupling aspheric lens, the lens focus at the waveguide output was optimized for the maximum coupling of the long-wavelength spectral features.

**GVD and generalized nonlinear Schrödinger equation (NLSE) calculations**

The generalized nonlinear Schrodinger's equation (gNLSE)[52] is used to model supercontinuum generation in the Si waveguides. The equation is commonly written as[38]

$$\frac{\partial E(z,\tau)}{\partial z} - i\sum_{k\geq 2} i^k \frac{\beta_k}{k!} \frac{\partial^k E}{\partial \tau^k} E(z,\tau) = -\frac{\alpha}{2}E(z,\tau) + \left(1 + \frac{i}{\omega_0}\frac{\partial}{\partial \tau}\right)\left(i\gamma|E|^2 E - \frac{\gamma_{3PA}}{3A_{eff}^2}|E|^4 E\right) - \frac{\sigma}{2}(1+i\mu)N_c E,$$

where $z$ is the propagation direction, $\tau = t - \beta_1 z$ is the temporal coordinate frame of reference of the pump pulse, $\alpha$ is the linear loss coefficient, $\gamma_{3PA}$ is the 3PA coefficient, $\sigma$ is the FCA cross-section, and $\mu$ is the FCD cross-section. $N_c$ is the free-carrier density in the waveguide, which evolves as $\frac{\partial N_c(z,\tau)}{\partial \tau} = -\frac{N_c}{\tau_{eff}} + \frac{\gamma_{3PA}}{3\hbar\omega_0 A_{eff}^3}|E(z,\tau)|^6$, where $\tau_{eff}$ is the free-carrier relaxation time.

We solved the gNLSE using the split-step Fourier method (SSFM) with an adaptive step-size algorithm using the following parameters: $n_2 = 6\times10^{-18} \frac{m^2}{W}$ (ref. 36), $\omega_0 \simeq 2\pi\times 98$ THz, $\gamma_{3PA} = 2\times10^{-26} \frac{m^3}{W^2}$ (ref. 53), and $\sigma = 5.65\times10^{-21}\, m^2$ (ref. 52). The FCD cross-section, $\mu = \frac{4\pi}{\lambda}\frac{1}{\sigma}\frac{n_0}{n_i}\left(\frac{8.8\times10^{-28}N_m + 8.5\times10^{-24}N_m^{0.8}}{N_m}\right)\left(\frac{\lambda}{1.55}\right)^2$ (ref. 52), with $N_m = \frac{\sqrt{\pi}\beta_{3PA}P_0^3 T_0}{3\sqrt{3}h\nu_0 A_{eff}^3}$, and $\frac{n_0}{n_i}$ are the maximum free-carrier induced charge density and the ratio between the bulk refractive index and effective mode index for Si, respectively. These are derived for 3PA-induced free carriers following the method explained in Ref. 52 and are evaluated for each waveguide independently with $P_0$ being the peak power in the waveguide, $T_0 = 100\, fs$ pulse duration, and $\nu_0 = \frac{\omega_0}{2\pi}$. The value for the linear loss was measured for each waveguide individually, and was roughly $\alpha \simeq 5$ dB/cm across the chip. The effective area of the mode, $A_{eff} \simeq 1\, \mu m^2$, was calculated via finite-element-method analysis for each waveguide individually. The free-carrier relaxation time is set to $\tau_{eff} = 1$ ns (ref. 37). The input pulse was measured to be a Gaussian pulse centered at $\lambda =$ 3.06 *μm* with a full-width half-max (FWHM) duration of 100 fs and an average power of 60 mW at the input facet of the waveguides.

**RIN measurement**

After collecting the waveguide output, the supercontinuum spectra are sent through a series of long-pass and band-pass filters. After the filters 500 nm FWHM mid-IR bands centered at 3.8 μm, 5 μm, and 5.4 μm was sent into a 200 MHz bandwidth (3 dB), amplified, liquid-nitrogen cooled Mercury-Cadmium-Tellurium (MCT) detector. The output of the detector was then monitored with a Vector Signal Analyzer to perform the RIN measurements. The spectra presented in Fig. 3 are measured in five frequency decades from 100 Hz to 10 MHz and then stitched together with each measurement averaged 10 times.

**Dual-Comb Spectroscopy**

The dual-comb spectroscopy setup consists of two mid-IR frequency combs operating at repetition rates of ~ 100 MHz. In comb 1, the mid-IR light is generated in a Si waveguide while in comb 2 a DFG comb, similar to the one described earlier, is constructed to generate mid-IR directly in the PPLN. This system is modified from the waveguide 3.06 μm pump system to allow tunable light generation from 2.6 μm to 5.2 μm. The amplified 1 μm beam in the pump branch has 2 W of average power. In the signal branch, the EDFA-amplified 1550 nm light was sent to a HNLF with zero-dispersion wavelength of 1520 nm for supercontinuum generation in the 1.3 μm to 1.6 μm range. The supercontinuum signal light is then combined and temporally overlapped with the 1.03 μm pump. In this setup, a 1 mm PPLN crystal was used for the DFG process with multiple poling periods enabling tunable idler generation in the 3 μm to 5 μm range, depending on the poling period. We operate the dual-comb systems in a wavelength band centered at 4.85 μm, with the repetition rates stabilized to get $\Delta f_{rep}$ = 512 Hz. A gas mix of 0.1% diluted OCS in 660 torr nitrogen is prepared in a 30 cm long gas cell. The beam from comb 2 is then passed through the cell and is later combined with the beam from comb 1. The resulting dual-comb heterodyne pulse train is then recorded with the amplified MCT detector. The output of the photodetector is monitored with an oscilloscope which is triggered at the center burst peak of the heterodyne interferogram to enable long averaging. The interferogram measurements are performed for two experimental schemes, namely, with and without the gas cell. The Fourier transform of the measured heterodyne signals give the optical spectrum of the dual-comb system with and without gas cell, respectively. We then use the optical spectrum without the cell as a reference to normalize the absorption spectrum.

Such a dual-comb system requires locking and stabilization of the lasers to keep the $\Delta f_{rep}$ at 512 Hz for averaging of the interferograms without losing the spectral features. Since our dual-comb system is based on two offset-free DFG lasers, we only need to lock the repetition rates to achieve the desired stability. A microwave circuit is designed and used for this purpose in which the residual 1 μm pump beams after the PPLN crystals are collected and sent into two fast photodetectors. A microwave band-pass-filter is placed at the output of the detectors. These filters pass the 100x harmonic of the repetition rates at ~ 9.969 GHz which we call 100f frequency in comb 1 and $100(f+\Delta f_{rep})$ in comb 2. The rest of the harmonics (including the repetition rate tone) are coupled to a side output of the band-pass filters. The repetition rate tone of the comb 1 laser is used to clock an RF frequency synthesizer. Two channels of the frequency synthesizer are used to generate RF frequencies of $f_0 + 100\Delta f_{set} = 500.0512$ MHz and $f_0 = 500$ MHz. These are then mixed with the 9.96 GHz harmonics of the comb 1 and comb 2 arms, respectively before they are passed through bandpass filters to only keep the added frequencies ($100f+f_0+100\Delta f_{set}$ in comb 1 and $100f+f_0+100\Delta f_{set}$ in comb 2). In the final stage of the circuit these two RF waves are mixed and band-pass filter to only keep the subtracted RF frequency of $100(\Delta f_{set} - \Delta f_{set})$. This signal is then sent to a loop filter module controlling the electro-optic modulators and piezo controllers in the comb 2 oscillator to stabilize the repetition rate difference at $\Delta f_{rep} = \Delta f_{set}$. In this scheme comb 1 laser acts as the master system with the comb 2 being the slave. Since the comb 1 repetition rate tone is used to clock the RF synthesizers, its phase noise is transferred to the comb 2 system causing the noises to cancel out when the $\Delta f_{rep}$ is referenced and stabilized.

**Acknowledgements**

This is a contribution of NIST, an agency of the US government, not subject to copyright. Product disclaimer: Any mention of commercial products is for information only; it does not imply recommendation or endorsement by NIST. This work was supported by NIST and the Defense Advanced Research Projects Agency (DARPA), Defense Sciences Office (DSO) under the SCOUT program. Si waveguide samples are fabricated in the NIST Boulder microfabrication facility.

**Author contributions**

N.N., D.A.W. and J.M.S. designed and fabricated the Si waveguides. N.N., D.L.M., F.C.C., H.T., and C.F. contributed to the optical measurements, noise characterization and spectroscopy experiments. N.N., A.K. and J.C. contributed the to the theoretical analysis with generalized nonlinear Schrodinger equation. The project was led, organized, and coordinated by S.W.N., R.P.M., J.M.S. and S.A.D. All authors contributed to the analysis of the data and writing of the manuscript.

**Additional Information**

Supplementary information is available in the online version of the paper. Reprints and permissions information is available online at www.nature.com/reprints. Correspondence and requests for materials should be addressed to N.N., R.P.M. and S.A.D.

**Competing financial interests**

The authors declare no competing financial interests.

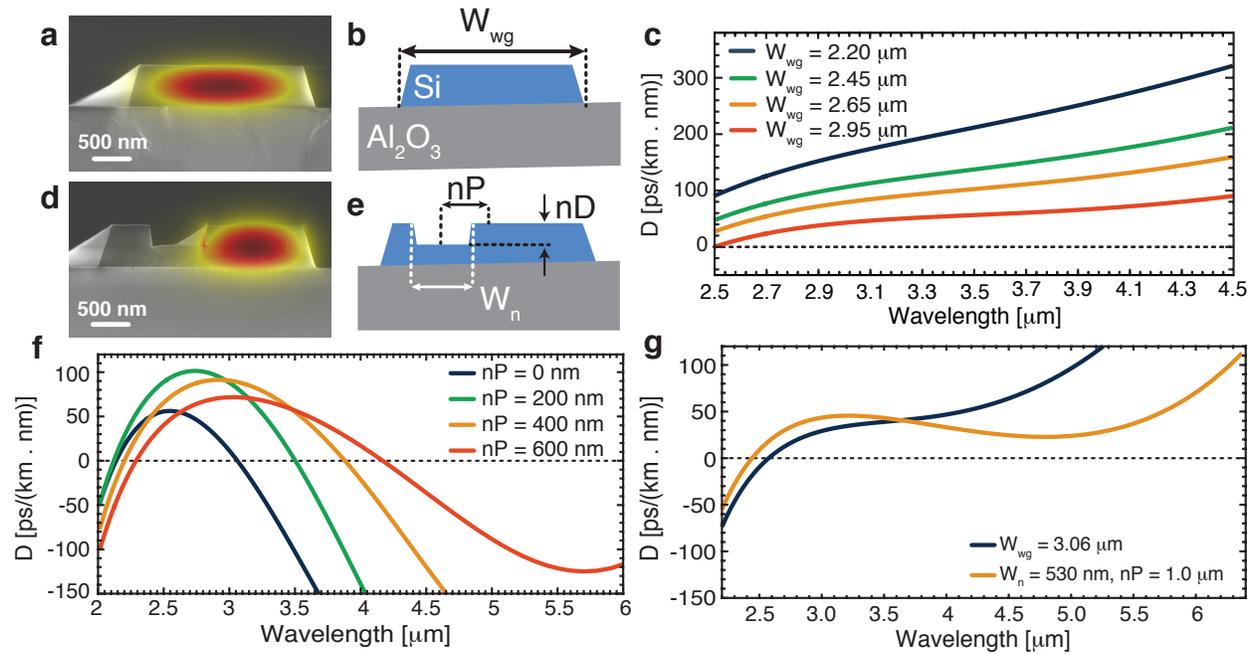

**Figure 1 | Waveguides, GVD designs and cross-section SEM micrographs. a, b, c)** SEM micrograph, cross-section diagram, and GVD profiles of fabricated strip waveguides, respectively. These waveguides are designed for multiband light generation in targeted 3 - 4.2 µm bands. **d, e, f)** SEM micrograph, cross-section diagram, and GVD curves of fabricated notch waveguides. The long-wavelength zero dispersion point can be tuned and pushed to longer wavelengths by increasing the notch position relative to the center of the waveguide, enabling efficient light generation up to 5.5 µm. The 3 µm input modes of the waveguides are overlaid on their SEM images while the waveguide width, notch width, notch depth, and notch position are labeled with $W_{wg}$, $W_n$, nD, and nP, respectively,

**g)** Designed GVDs of two waveguides for octave-span supercontinuum generation. The dark blue curve is for a 3.06 μm wide strip waveguide, while the orange curve represents the GVD of a notch waveguide with $W_n$ = 530 nm and, nP = 1 μm.

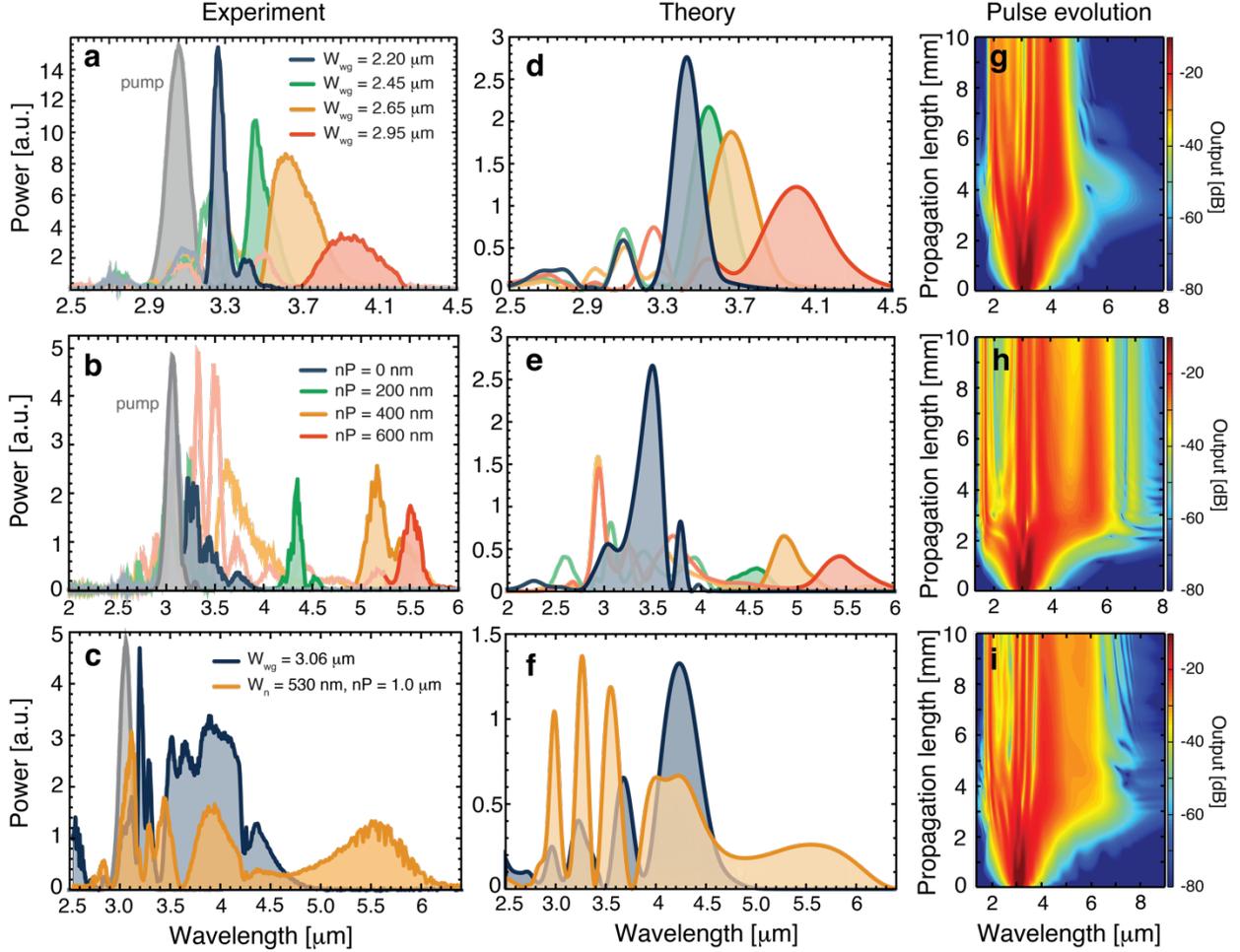

**Figure 2 | Targeted band supercontinuum generation in Si waveguides. a, b)** Measured output spectra of strip and notch waveguides, respectively. **c)** Octave span broadening in mid-IR extending to 6.2 μm from a strip (blue curve) and notch (orange curve) waveguide **d, e, f)** Generalized nonlinear Schrödinger equation calculated supercontinuum spectra of the waveguide devices under investigation for strip waveguides, notch waveguides, and octave-span broadening devices, respectively. **g, h , i)** Calculated two-dimensional frequency domain, pulse-evolution plots along the length of the waveguides. Data is presented for the broadest spectra in (d, e, and f). Experimental data are presented in arbitrary units due to lack of absolute power information when acquiring infrared spectrum with an FTIR.

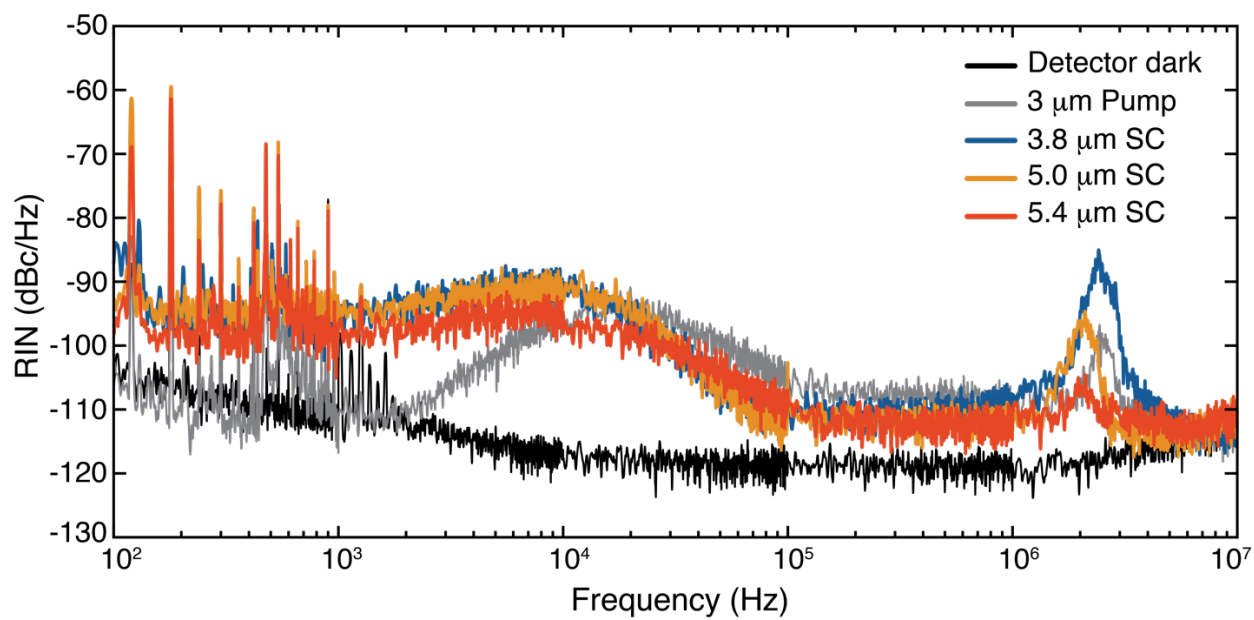

**Figure 3 | Measured relative-intensity-noise (RIN) of the generated mid-IR bands.** The RIN data is measured for the mid-IR bands of 3.8 μm, 5.0 μm, and 5.4 μm at the output of the Si waveguides and compared with that of the 3 μm pump. The detector dark measurement is presented in black

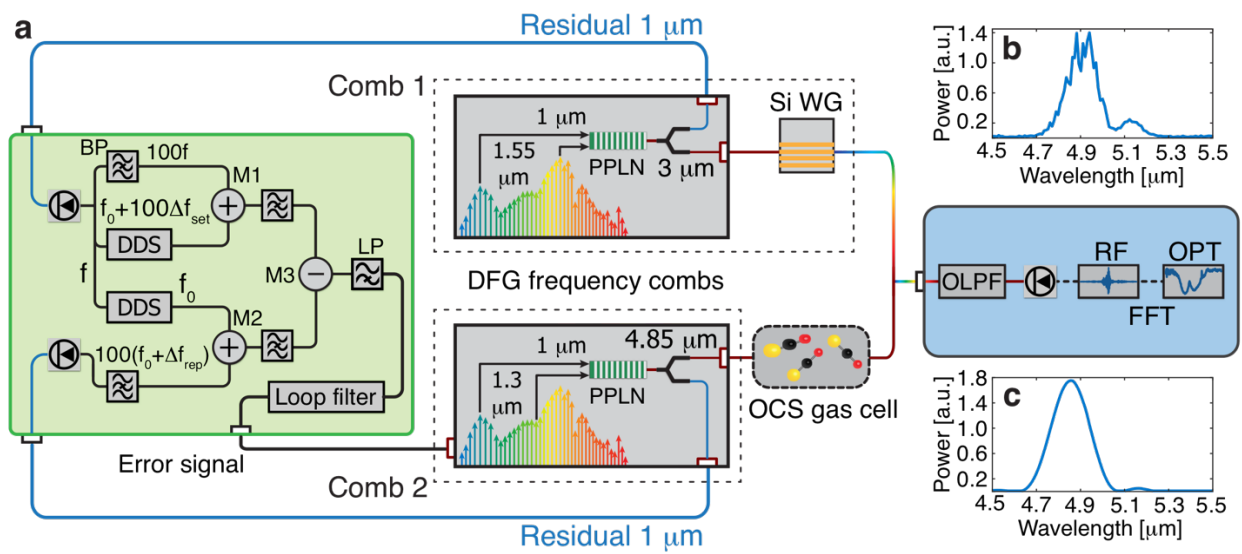

**Figure 4 | Constructed Si-chip based dual-comb setup for mid-IR spectroscopy at 4.85 μm. a)** Diagram of the constructed Si-chip-based dual-comb spectroscopy setup. RF: Radio frequency, DDS: Direct digital synthesizer, BP: microwave band-pass filter, M1, M2, M3: microwave mixers, LP: microwave low-pass filter, PPLN: periodically-poled-lithium-niobate, DFG: Difference frequency generation, WG: waveguide, OLPF: Optical long-pass filter,

FFT: fast-fourier-transform, OPT: Optical domain spectrum. **b)** Spectrum of the dispersive wave generated in the Si waveguide and used in dual-comb spectroscopy of OCS with **c)** the spectrum generated from the DFG in PPLN in *"Comb 2"*.

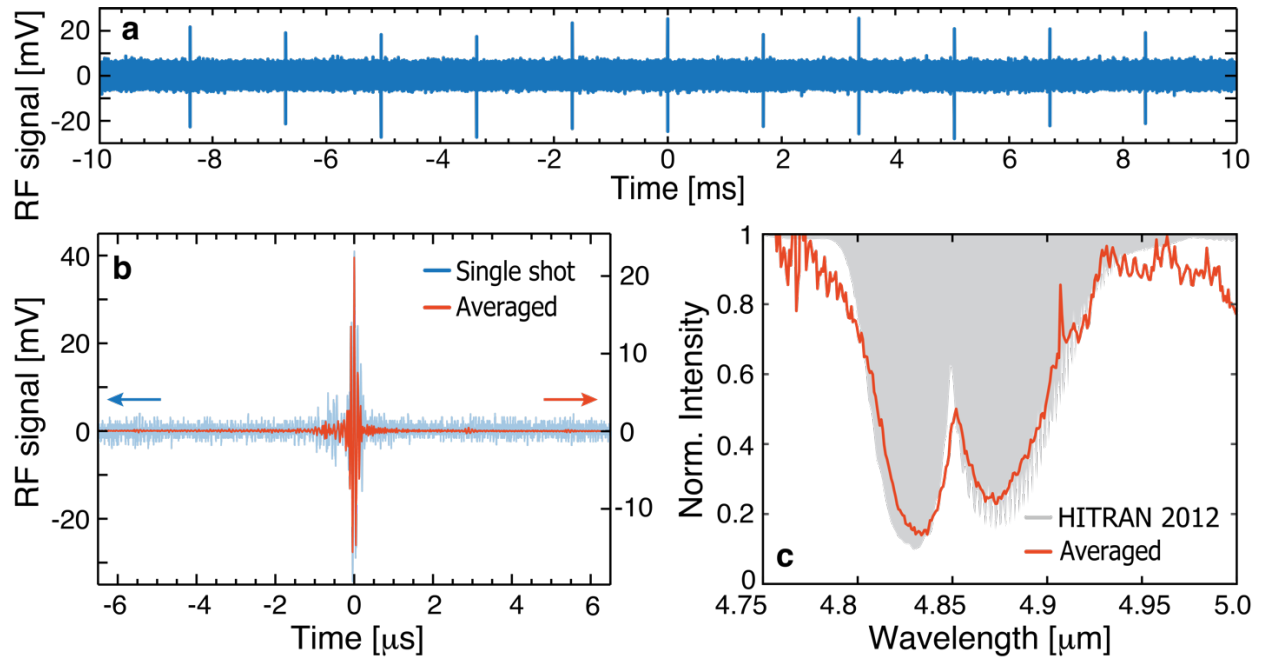

**Figure 5 | Dual-comb spectra of OCS absorption in 4.8 μm mid-IR band. a)** Time domain, single shot, pulse train of the dual-comb interferogram in 20 ms time span. **b)** The single shot (light-blue) and 16384 times averaged (red) center burst of the dual-comb heterodyne signal along with the molecular free-induction-decay from OCS sample in a zoom in 12 μs time span. The single shot signal is plotted on the left vertical axis with the averaged data ploted on the right. **c)** The normalized absorption spectrum of OCS is presented at 16384-times averaging and compared with the reference spectrum from HITRAN 2012 database. The normalized spectra in (**c**) are acquired by calculating the Fourier transform of the heterodyne signal in (**c**). The absorption spectra are normalized to that of the dual-comb system without gas cell.